\documentclass[sigconf]{acmart}

\usepackage{booktabs} 

\setcopyright{rightsretained}

\begin{document}
\title{Nonlinear Dance Motion Analysis and Motion Editing using Hilbert-Huang Transform}

\author{Ran Dong}
\affiliation{%
	\institution{University of Tsukuba}
	\city{Tsukuba} 
	\state{Ibaraki, Japan} 
	\postcode{305-8577}
}
\email{dong@cavelab.cs.tsukuba.ac.jp}

\author{Dongsheng Cai}
\affiliation{%
	\institution{University of Tsukuba}
	\city{Tsukuba} 
	\state{Ibaraki, Japan} 
	\postcode{305-8577}
}
\email{cai@cs.tsukuba.ac.jp}

\author{Nobuyoshi Asai}
\affiliation{%
	\institution{University of Aizu}
	\city{Aizu} 
	\state{Fukushima, Japan} 
	\postcode{965-8580}
}
\email{nasai@u-aizu.ac.jp}

\begin{abstract}
Human motions (especially dance motions) are very noisy, and it is hard to analyze and edit the motions. To resolve this problem, we propose a new method to decompose and modify the motions using the Hilbert-Huang transform (HHT). First, HHT decomposes a chromatic signal into "monochromatic" signals that are the so-called Intrinsic Mode Functions (IMFs) using an Empirical Mode Decomposition (EMD) \cite{huang2014hilbert}.  After applying the Hilbert Transform to each IMF, the instantaneous frequencies of the "monochromatic" signals can be obtained. The HHT has the advantage to analyze non-stationary and nonlinear signals such as human-joint-motions over FFT or Wavelet transform.

In the present paper, we propose a new framework to analyze and extract some new features from a famous Japanese threesome pop singer group called "Perfume", and compare it with Waltz and Salsa dance.  Using the EMD, their dance motions can be decomposed into motion (choreographic) primitives or IMFs. Therefore we can scale, combine, subtract, exchange, and modify those IMFs, and can blend them into new dance motions self-consistently. Our analysis and framework can lead to a motion editing and blending method to create a new dance motion from different dance motions. 

\end{abstract}

%
%


\copyrightyear{2017}
\acmYear{2017}
\setcopyright{acmcopyright}
\acmConference{CGI '17}{June 27-30, 2017}{Yokohama, Japan}\acmPrice{15.00}\acmDOI{10.1145/3095140.3095175}
\acmISBN{978-1-4503-5228-4/17/06}

\settopmatter{printacmref=false} 


\begin{teaserfigure}
	\includegraphics[width=\textwidth]{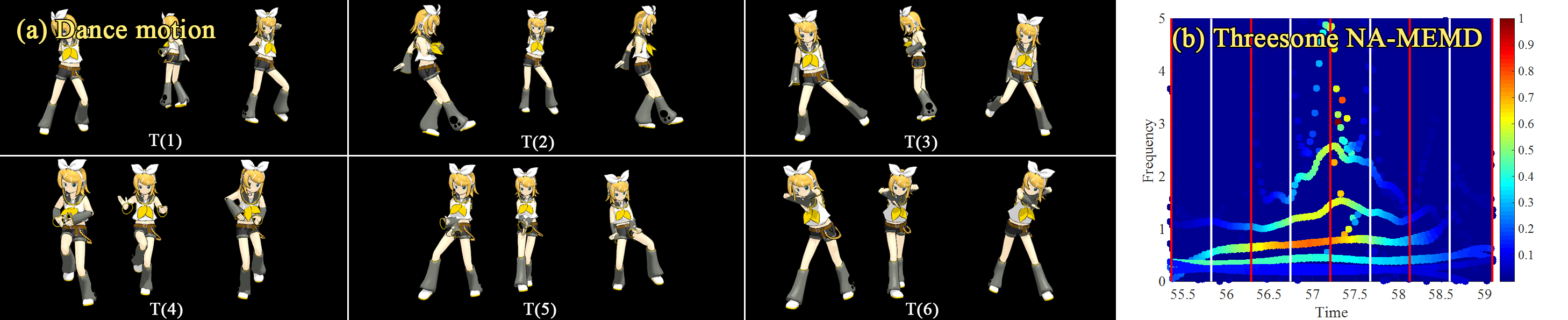}
	\caption{(a) Six-time shots of a four-second Japanese threesome pop unit "Perfume" dance motions.  (b) Hilbert spectrum (Energy (color)-Frequency (y-axis)-Time (x-axis)) of the dance motion data. The white and red lines are weak and strong beats, respectively.}
	\label{fig:teaser}
\end{teaserfigure}

\begin{teaserfigure}
	\includegraphics[width=\textwidth]{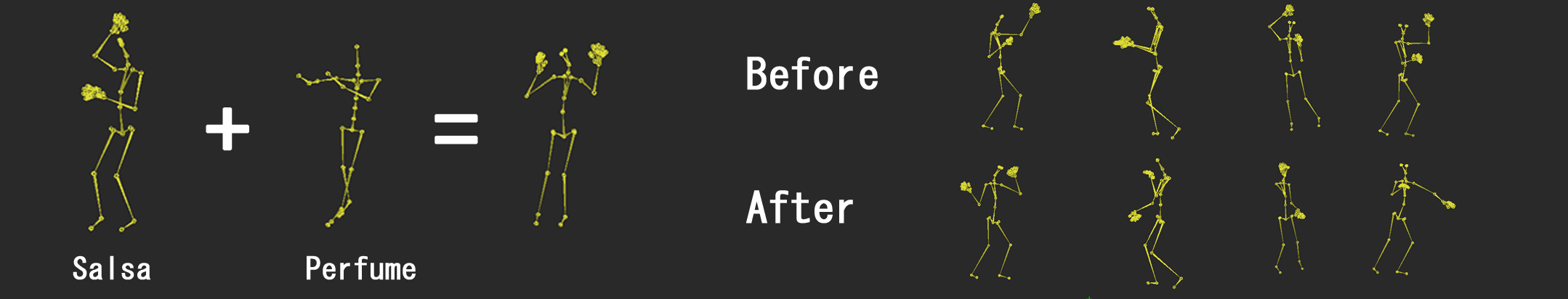}
	\caption{Salsa dance motion with Japanese threesome pop unit "Perfume" choreographies.  Decomposing Perfume and Salsa dance into different distinct modes (IMFs) using NA-MEMD. Perfume upper body motion IMFs are blended with the Salsa motion IMFs. A new Perfume dance movements with Salsa steps are created.}
	\label{fig:teaser}
\end{teaserfigure}

\maketitle

\section{Introduction}

We analyze and decompose a four-second threesome techno pop dance by Perfume as shown in Fig. 1. The threesome dancers' motions are almost synchronous and occasionally asynchronous. Applying the Noise-Assisted Multivariate Empirical Mode Decomposition (NA-MEMD) to all three dancers' joint angle data, the hip movements of one dancer are now clearly decomposed into five different almost harmonic IMFs as shown in the left panel (b) of Fig. 1 (Hilbert spectrum plot). In the center of the plot, at t=57.25 sec., two modes peaked at 2.2, and 5 Hz indicate the left and right leg rotational motions respectively, and now this four-second noisy motion data is clearly decomposed into five modes or IMFs and one residual motion entitled "trend". Note that these clear left and right leg rotations cannot be detected using FFT and wavelet analyses. When we apply the same NA-MEMD to Waltz, and Salsa dances, these dance motions also can be apparently decomposed into 9-11 different IMFs and one trend. 

Appropriately using the NA-MEMD, dance motions can be completely decomposed into different IMFs that correspond to different nonlinear "monochromatic" or "choreographic" motions, and one trend that is "non-choreographic" motions. Consequently, these different "choreographic" motions or corresponding IMFs can be cut, pasted, eliminated, subtracted, added, scaled, and they also can be edited and blended with the different dance motions or corresponding IMFs smoothly. In Fig. 2, we replace the Salsa upper body IMFs with those of Perfume. Thus, Perfume dance can easily be converted into that with Salsa steps.  

In this paper, first, we introduce previous works and review HHT briefly. Second, using HHT, we propose a new method and framework to analyze and edit dance motions. Third, using our proposed framework, we analyze a few dances (mainly Perfume dance) and adapt and blend those dances to create a new dance with different tastes. Finally, we conclude our paper.

\section{Related Works}
Various motion blending techniques based on motion capture data are proposed for character animation. For examples, Heck suggests a method using parametric motion graphs to generate and interpolate a new motion between some basic motion data \cite{heck2007parametric}. Lee proposed a method editing existing human-like character motion data to obtain desired human actions \cite{lee1999hierarchical}. However, previous researches focus only on interpolating motions, and they cannot blend different movements into new motions. 

Unuma used Fourier coefficients to extract and edit motion capture data \cite{unuma1995fourier}. However, it is only possible to do so when animators have already known the characteristics of the targeted motions. It is inappropriate to use this method to edit dance motions composed of various unknown features.  

To resolve this problem, we propose a framework using HHT to decompose dance motion into nonlinear signals (monochromatic waves) to extract their features.  We also offer a method to edit and blend different dances to generate new dance motions using HHT.

\section{Hilbert-Huang Transform}
\subsection{Hilbert Transform}

\begin{figure}[h]
	\includegraphics[width=3.0in]{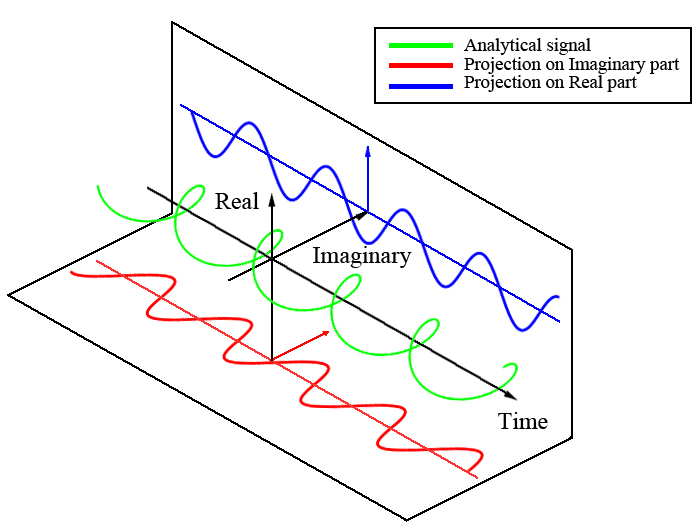}
	\caption{Analytical signals with a real and imaginary part. They are quadrature functions (cited from \cite{bracewell1986fourier}).}
\end{figure}

An analytical signal $z(t)=z_{r}(t)+iz_{i}(t)$ varying in time is plotted in Fig. 3. The instantaneous amplitude and frequency of Fig. 3 can be expressed as follows:

\begin{equation}
\label{eqn:02}
A(t)=\sqrt{z_{r}^2(t)+z_{i}^2(t)} \qquad
\omega_0(t)=\frac{d}{dt}tan^{-1}\frac{z_{i}(t)}{z_{r}(t)} 
\end{equation}

It is known that the Hilbert Transform (HT) converts the real part of the analytical function into its imaginary part. However, the Hilbert Transform assumes the signal is a monochromatic wave, and the real part of the analytical function can be expressed as $x(t)=A(t)cos(\omega_0 (t)t)$ as shown in Fig. 3. Thus, Hilbert Transform can be defined as follows:

\begin{equation}
\label{eqn:01}
z_{i}(t)=y(t)=\frac{1}{\pi}PV\int_{-\infty}^{\infty} \frac{x(\tau)}{t-\tau}\,d\tau=\frac{1}{{\pi}t}*x(t)
\end{equation}
In Eq. (2), the integration means the Cauchy principal value (PV) integration \cite{bracewell1986fourier}.

Since the Hilbert Transform (HT) assumes the signal is a monochromatic wave $x(t)$, it is not suitable to apply HT to the chromatic signals that are true for many real-world signals. Thus, Huang proposed a method entitled the Empirical Mode Decomposition (EMD) to decompose the chromatic signals into a set of monochromatic signals called the Intrinsic Mode Functions (IMFs). 

The essential part of the HHT is the "Empirical Mode Decomposition (EMD)". A single-variate or multivariate chromatic signal can be decomposed into a finite set of "Intrinsic Mode Functions (IMF)" and a residual called "trend" \cite{huang2014hilbert}. After applying motion captured joint angle data into the Noise-Assisted Multivariate Empirical Mode Decomposition (NA-MEMD) \cite{ur2013emd}, we can decompose dancers' "choreographic" motion primitives into separate IMFs and a trend that corresponds to dancers' postures including transitions in time.

\subsection{Empirical Mode Decomposition (EMD)}

In this section, we briefly introduce the EMD algorithm and the IMF definition (We do not explain the details of IMF and its residuals $r$ in the present paper.  For the details, please see \cite{huang2014hilbert}).  The Hilbert Transform can be applied to the signals after decomposing the original signal into an IMF set and a "trend" \cite{huang2014hilbert}.

\subsection{EMD Algorithm}

Intrinsic Mode Function (IMF) is defined as follow \cite{huang2014hilbert}: 
\begin{itemize}
\item The number of extreme value is equal to that of zero-crossings, or the difference is 1.

\item The average of maximum and minimum envelopes is 0 at any time.
\end{itemize}

{\noindent}Here is the EMD algorithm \cite{huang2014hilbert}:

\begin{enumerate}
\item  Calculate residual $r(t)$  (Let $r(t)=x(t)$ in the first time) as follows:

\begin{equation}
\label{eqn:01}
r(t)=x(t)-\sum_n C_n(t)
\end{equation}

\item  Initialize $c(t)=r(t)$  and extract the Intrinsic Mode Function (IMF)

	\begin{enumerate}
	\item Find maximum envelope $u(t)$ and minimum envelope $l(t)$ of $c(t)$
	
	\item Subtract the average envelope from $c(t)$
	
	\begin{equation}
	\label{eqn:01}
	c_{new}(t)=c_{old}(t)-\frac{u(t)+l(t)}{2}
	\end{equation}
	
	\item If the convergence condition $SD$ (0.2-0.3) is satisfied, add $c(t)$ into the IMF set.
	\begin{equation}
	\label{eqn:01}
	SD=\sum_n \frac{(c_{old}(t)-c_{new}(t))^2}{{c_{old}}^2(t)}
	\end{equation}
	\end{enumerate}
\item  Repeat step 1 and 2 to extracts all IMFs. \\

\end{enumerate}

\subsection{Multivariate EMD (MEMD)}
For multi-channel or multivariate analytical signals, it is impossible to use the EMD to decompose "chromatic" signals. The Multivariate EMD (MEMD) is proposed for multi-channel or multivariate signals such as dance motion data that are composed of all joint angles of a human body \cite{rehman2009multivariate}.

Due to MEMD Filter bank function that can remove white noises of the data, in NA-MEMD, Gaussian White Noises (GWN) is proposed to be packed into one extra data channel.   Thus, the mode mixing that causes the HHT inaccurate can be reduced or eliminated \cite{ur2013emd}. 

\begin{figure}[h]
	\includegraphics[width=3.5in]{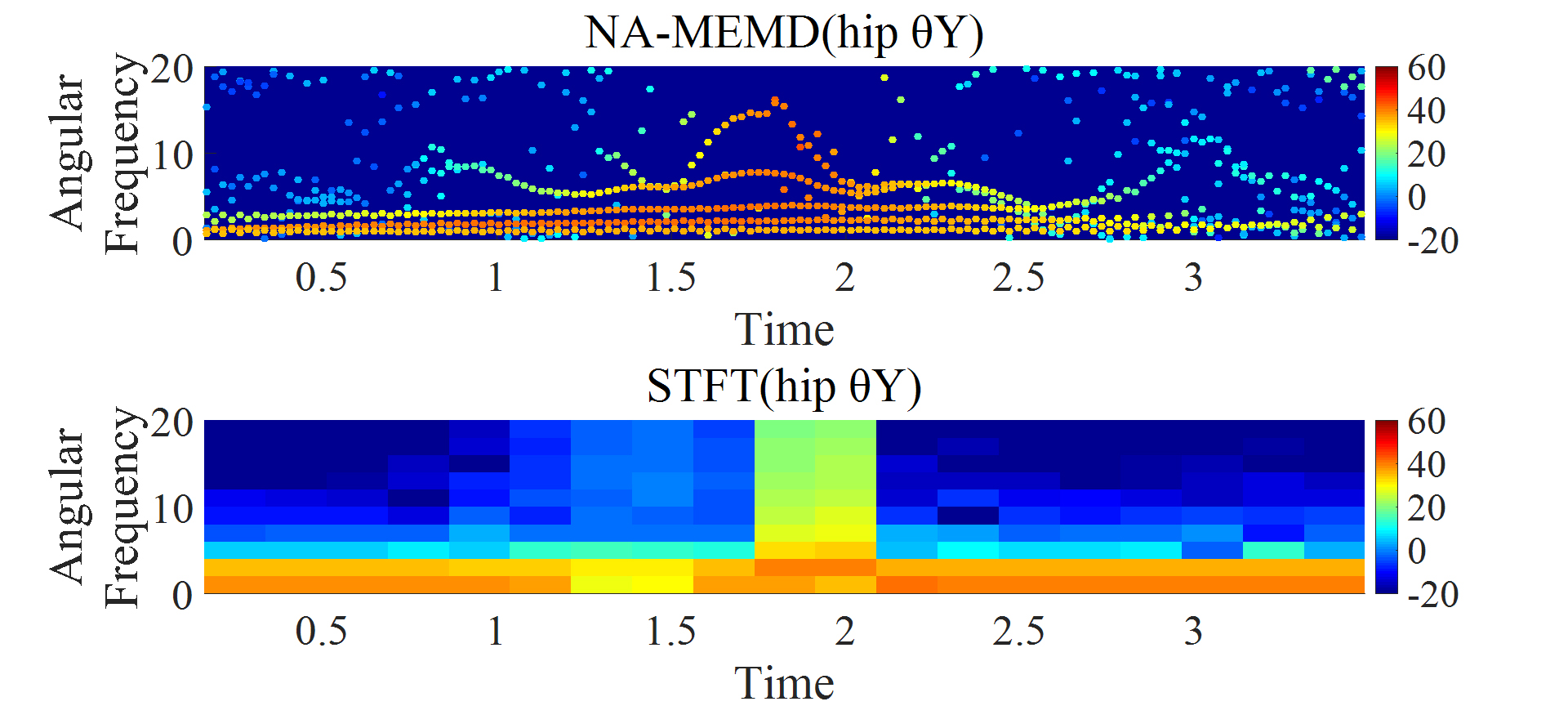}
	\caption{A four-sceond motion comparisons between HHT and Short-Time-Fourier-Transform (STFT). The motion is the same as that of Fig. 1.}
\end{figure}

As shown in Fig. 4, HHT can accurately capture five different choreographic motion primitives or IMFs of the left leg and right leg, while the Short-Time-Fourier-transform (STFT) only can capture vague high-speed motions around the center.  It is desirable to observe dance movements in a frequency domain, and HHT certainly has some advantages over STFT, because the uncertainty principles limit the STFT resolutions. 

\section{Proposed Framework of Analyzing and Decomposing Dance Motions}
\subsection{Dance Motions Analysis Framework}
Fig. 5 shows our proposed framework using both motion capture data and music data. 
\begin{figure}[h]
	\includegraphics[width=3in]{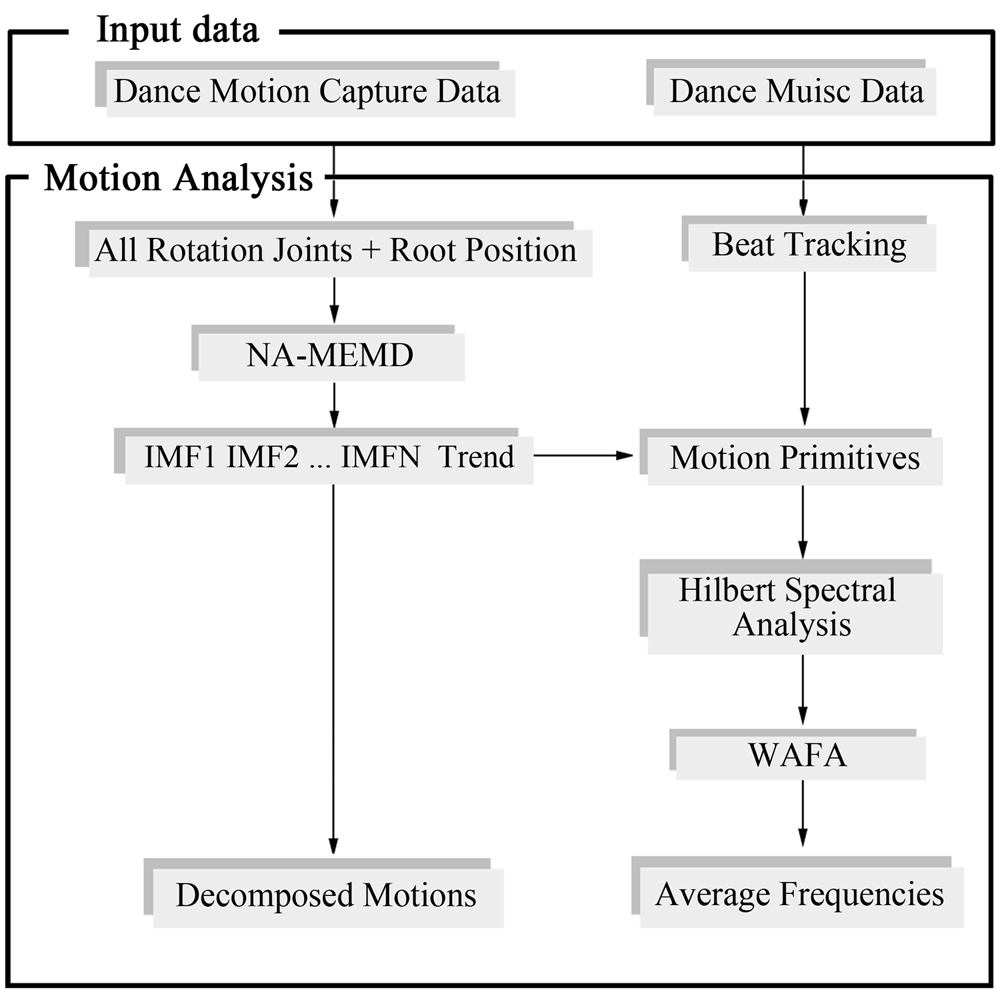}
	\caption{Proposed framework to decompose "choreographic" motion primitives and analyze dance motions using HHT.}
\end{figure}
\begin{description}
\item[Beat Tracking]
We used a beat tracking system suggested by Dan Ellis to segment the dance motions \cite{ellis2007beat}. Our system assumes a regular tempo based on a fixed beat per minute (BPM). Then, we calculate the relevancy between beats and music to extract the beat itself. Please see the details in \cite{ellis2007beat}. 

\item[Motion Primitives]
To analyze and edit dance motions, it is necessary to segment dance motions into a set of short meaningful motions ("choreography"). A motion primitive is the most basic motion unit in human motions, and it can be expressed as a time series of key poses. To analyze dance motions and extract motion primitives, first, we must track beats in dance music in our system because each primitive is synchronizing with the beats. To edit the dance motion primitives, the motions have to be segmented based on the beats.

\item[NA-MEMD]
The motion-captured human motions are composed of many body joints with the Eulerian angles ${\theta}x$,${\theta}y$,${\theta}z$. To handle these data using the EMD, we generate the n-dimensional envelopes by projecting different joint signals along different directions in the n-dimensional space using the n-dimensional sphere. After we obtain the n-dimensional envelope, IMFs are obtained using the same algorithm, as introduced in subsections 3.3 and 3.4 \cite{rehman2009multivariate}. Usually, motion captured kinematic data are extremely noisy.  The NA-MEMD can significantly reduce the noises of the motion-captured data to obtain their IMFs as shown in Fig. 1(b) \cite{ur2013emd}.

\item[Weighted Average Frequency Algorithm]
Since we obtain the instantaneous frequency and amplitude in Eq. (1) using HHT, averaging them is necessary.  It is because that the instantaneous frequency and amplitude in one flame are meaningless for one dance movement due to the nature of the instantaneous frequency and amplitude. So we have to observe and define the frequency in one smallest time unit i.e. one motion primitive time unit. 
The Weighted Average Frequency Algorithm (WAFA) is a method to average IMFs to obtain meaningful data.  Besides, the real signals like motion captured data are always very noisy \cite{niu2012weighted}. Applying WAFA to one motion primitive time length, the meaningful and denoised IMFs can be obtained.
\end{description}

\subsection{Limitations of the proposed framework}
EMD is a decomposition technique without a mathematical proof. Therefore, there is no guarantee that dance motions can be decomposed into proper IMFs. Our proposed framework may have the following limitations:

\begin{description}
\item [An Over-Decomposed IMF] 
In an EMD, a chromatic signal is decomposed into a set of "monochromatic" signals from high to low-frequency signals. However, dance motions are not exactly choreographed by monochromatic motions. Thus, one motion primitives can be over-decomposed into a few similar motion IMFs. To avoid over-decompositions, we have to combine over-decomposed IMFs with a few adjacent IMFs manually.

\item [A Singular IMF]
The frequencies of decomposed IMFs decrease gradually with the decomposition order. However, a singular IMF is an IMF with an outlier frequency value \cite{huang2013threshold}. Usually, the singular IMFs can be observed in a low-frequency dance motion, when taking their WAFA frequencies. Sometimes, the singular IMF is not associated with its original dance primitives physically. To avoid this singularity, it is necessary to use a proper noise in NA-MEMD. In general, it is ideal that adding GWN whose amplitude is about 8\%-10\% of the original signal into NA-MEMD. It can avoid the mode mixing and the singular IMF \cite{ur2013emd}.

\end{description}
	
\section{Results}
\subsection{Dance Motion Analysis}
The threesome techno pop dance unit Perfume has been performed for more than ten years and considered to be matured and experienced dance group. Perfumes' dance is considered to be extremely difficult. However, at a glance, it seems catchy, simple and easy to dance. According to Perfume choreographer Mikiko, she expressed that the image of Perfume music is inorganic and closer to near future than the modern style. As a result, their dances have been arranged to express a puppet to obtain unworldly movements. 

The motion captured Perfume dance is compared with other motion-captured dances such as Waltz, and Salsa using the proposed HHT method.  Here, Table 1 shows the lengths and frame rates of four different dance motion captured data. 

\begin{table}[h]
	\caption{Motion Captured Data for Analysis}
	\label{tab:freq}
	\begin{tabular}{cccl}
		\toprule
		Dance Style&Perfume&Waltzy&Salsa\\
		\midrule
		Time(s) & 60.5& 52.6 & 28.5	\\
		Frame rate & 40& 30&40\\
		\bottomrule
	\end{tabular}
\end{table}

We apply the NA-MEMD to three hip Euler angles ${\theta}x$,${\theta}y$,${\theta}z$ for full dance motions to compare their features. Figures 6-8 show HHT spectrum of the one hip Eulerian angle for three different dances respectively.  Here, the angle of hip ${\theta}x$ for Perfume, Waltz and ${\theta}z$ for Salsa is plotted because they are the most significant angles in the dances. The beats are not displayed to avoid being overwhelmed. Each IMF average frequencies that are averaged for all dance time are listed in the left column of the figures.

\begin{figure}[h]
	\includegraphics[width=3.3in]{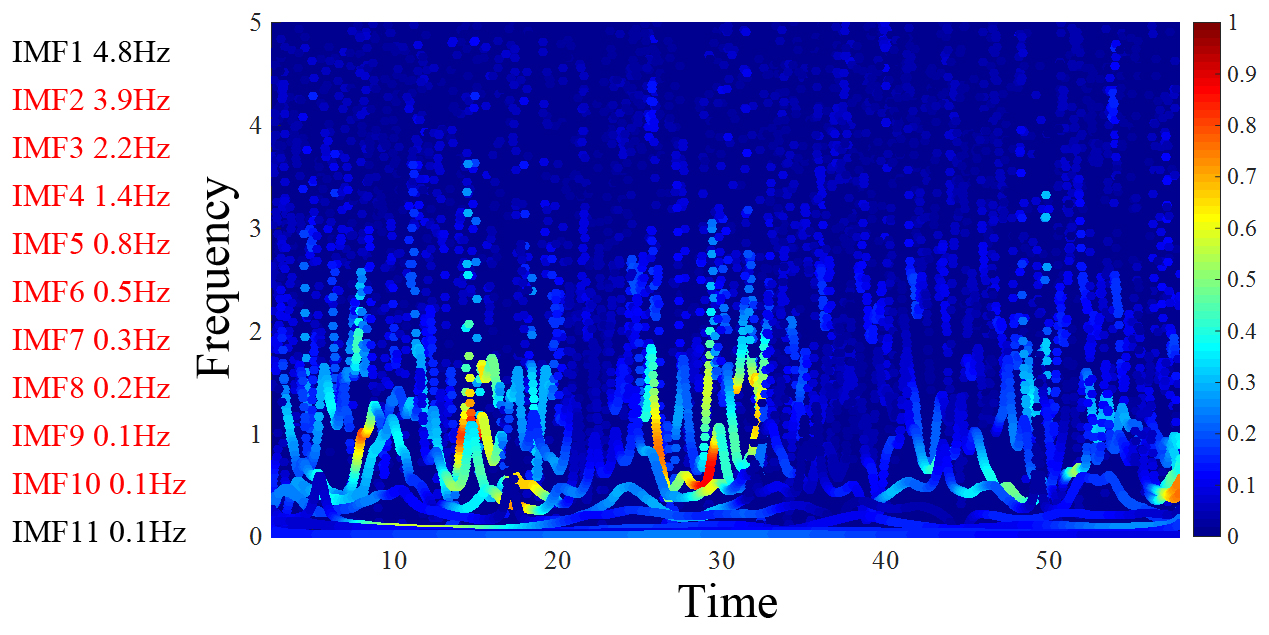}
	\caption{NA-MEMD Hilbert Spectrum of Perfume (hip  ${\theta}x$).}
\end{figure}

\begin{figure}[h]
	\includegraphics[width=3.3in]{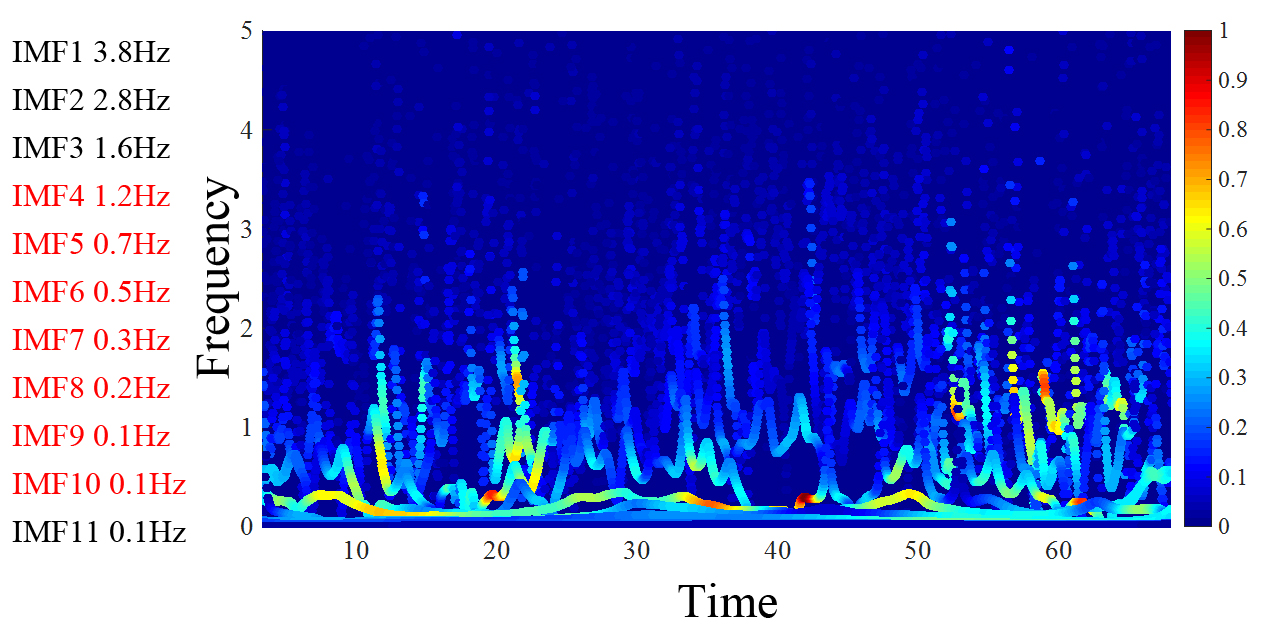}
	\caption{NA-MEMD Hilbert Spectrum of Waltz (hip  ${\theta}x$).}
\end{figure}

\begin{figure}[h]
	\includegraphics[width=3.3in]{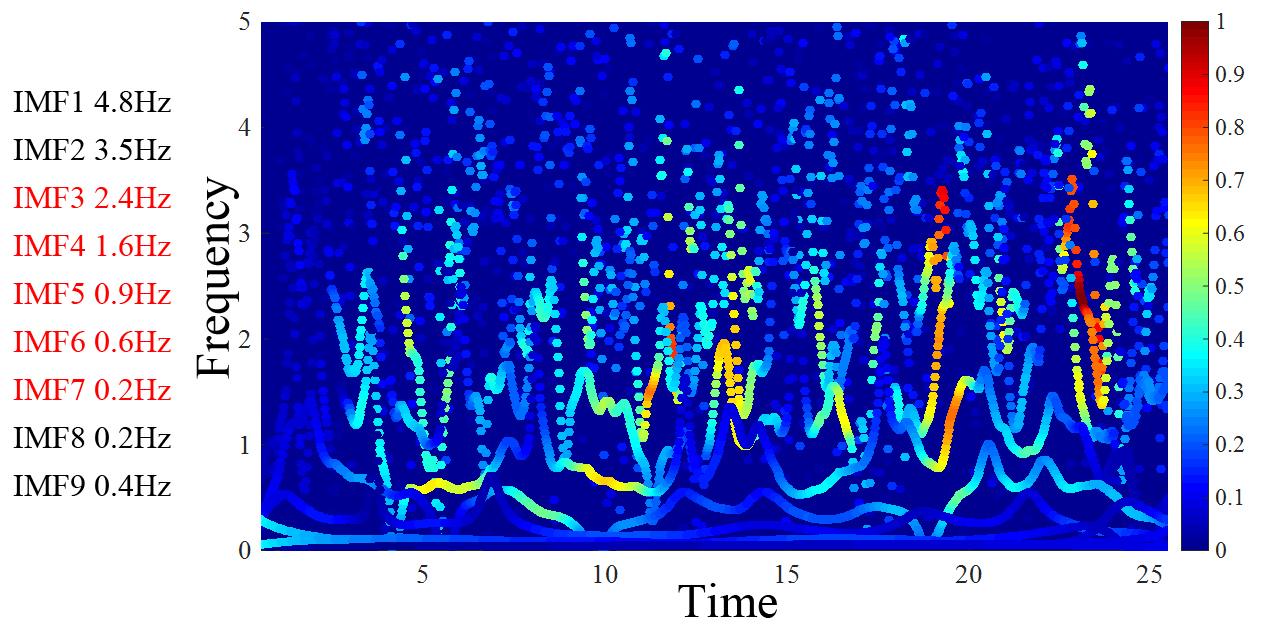}
	\caption{NA-MEMD Hilbert Spectrum of Salsa (hip ${\theta}z$).}
\end{figure}

In Figs. 6, 7, and 8, the Perfume, Waltz, and Salsa dance motions are respectively decomposed into 11, 11, and 9 IMFs ranging from 0.1 to 4.8Hz, from 0.1 to 3.8Hz,  and from 0.4 to 4.8Hz. The Salsa dance Hilbert spectrum shows IMF frequencies are noisy and concentrated around 2-4Hz. These results indicate that the Salsa motion is very up-tempo and the dance style is rather free, unorganized, and noisy. 

From our beat tracking analysis, Perfume, Waltz, and Salsa dance motion BPMs (Beat Per Minutes) are, respectively, 130, 84-93, and 150 (we estimate it as a slow type Salsa) \cite{Tempo}. The BPM of Perfume is about 130 and is as fast as that of Salsa. In addition, the highest WAFA frequency of IMF (IMF 1) of both Perfume and Salsa is the same and is 4.8 Hz, and much greater than the Waltz.  The most distinct Hilbert power spectrum of both Perfume and Salsa is located around 1-2 Hz, and higher than the Waltz of which the distinct Hilbert power spectrum are located around 0-1 Hz.  

On the other hand, the IMF number of Perfume is 11 and is as many as that of Waltz which is a rather slow tempo and classical dance. The Waltz is considered to have many distinct choreographic primitives, and each IMF should correspond to different choreography or motion primitives (The way how the different dance motions correspond to the different IMFs are not shown here due to the limit of the paper length). 

One of the most interesting features of the HHT analysis applied to the dance motions can be analyzed from the WAFA frequencies of IMFs listed as the red numbers in the left columns in Figs. 6-8. The WAFA frequencies are averaged over for whole dance time. From the bottom to the top of those red numbers, we can quickly to find the relation as follows: frequency of IMF (n) = frequency of IMF (n+1) + frequency of IMF (n+2). From the bottom to top, they form the Fibonacci sequences.  In Fig. 6 of Perfume dance, from the bottom to top, 0.1+0.1=0.2, 0.1+0.2=0.3, 0.2+0.3=0.5Hz..., etc., and except the top two and bottom IMFs, they all form Fibonacci sequences.   These Fibonacci relations indicate the dancers are moving their body strictly in a regular tempo and rhythms. Comparing the Perfume (Fig. 6) and Salsa (Fig. 8), the Perfume dancer follows more strictly Fibonacci relation than the Salsa dance.  

Following the Hilbert-Huang spectrum analysis, we can summarize the Perfume dance as follows: 
\begin{enumerate}
\item the Perfume dance IMF spectrum have as high frequency as those of up-tempo Salsa. 
\item the Perfume dance has as many IMFs as those of the traditional Waltz
\item Perfume dance average IMF frequencies follows more Fibonacci relations than the up-tempo Salsa. 
\end{enumerate}

We can assume the number of dance choreography is proportional to the IMF number. We know the Salsa is up-tempo and freestyle dance. The Perfume dance is as up-tempo as the Salsa but has as much kind of IMF or motion primitives as that of Waltz.   Also, Perfume dance motion is strictly in regular tempo and rhythm. These may explain the reason why people comment "Perfume's dance is visually simple but difficult to dance", and the choreographer Mikiko commented her dance as "puppet-like".

\subsection{Dance Motion Editing}

\begin{figure}[h]
	\includegraphics[width=3.2in]{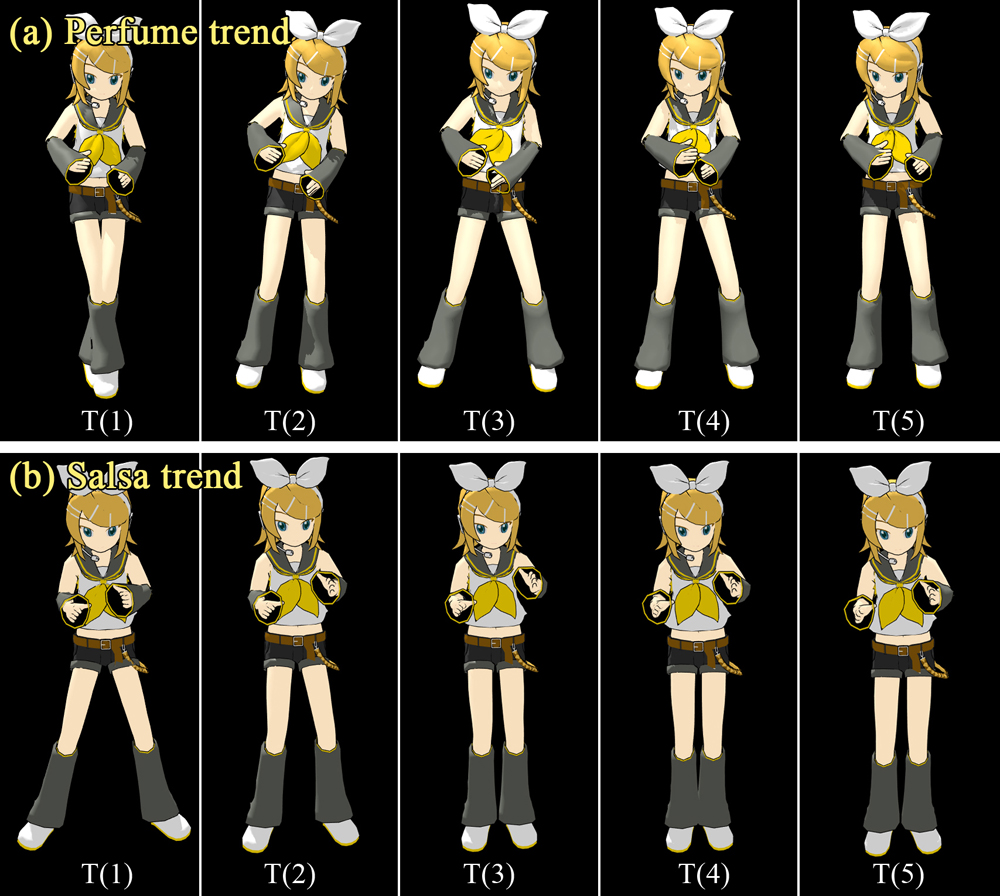}
	\caption{(a) Perfume trend, and (b) Salsa trend extracted from 55-second motion data by NA-MEMD.}
\end{figure}

The EMD can decompose chromatic signals into a finite set of monochromatic signals entitled IMFs and its residual entitled Trend. Applying the NA-MEMD to dance motion captured data, the different IMFs can be associated with the different motion primitives or choreographic motions, and the trend can be associated with the posture of the dance. We show dance motions by using free software MikuMikudance. Fig. 9 shows Perfume and Salsa trends, these trends have almost no distinct body movements and only show the gradual time transitions of their postures. In the figure, a significant difference between Perfume and Salsa trends can be found, especially, on the upper and lower bodies.

Because the EMD can decompose the motions into the distinct different motion primitives as "monochromatic" signals and trend, we can cut, paste, add, subtract, and scale the IMFs and trend appropriately, to produce new dance motions consistently. Also, we can add, blend, exchange IMFs from different dances and create a new dance with different tastes.  

\begin{figure}[h]
	\includegraphics[width=3.2in]{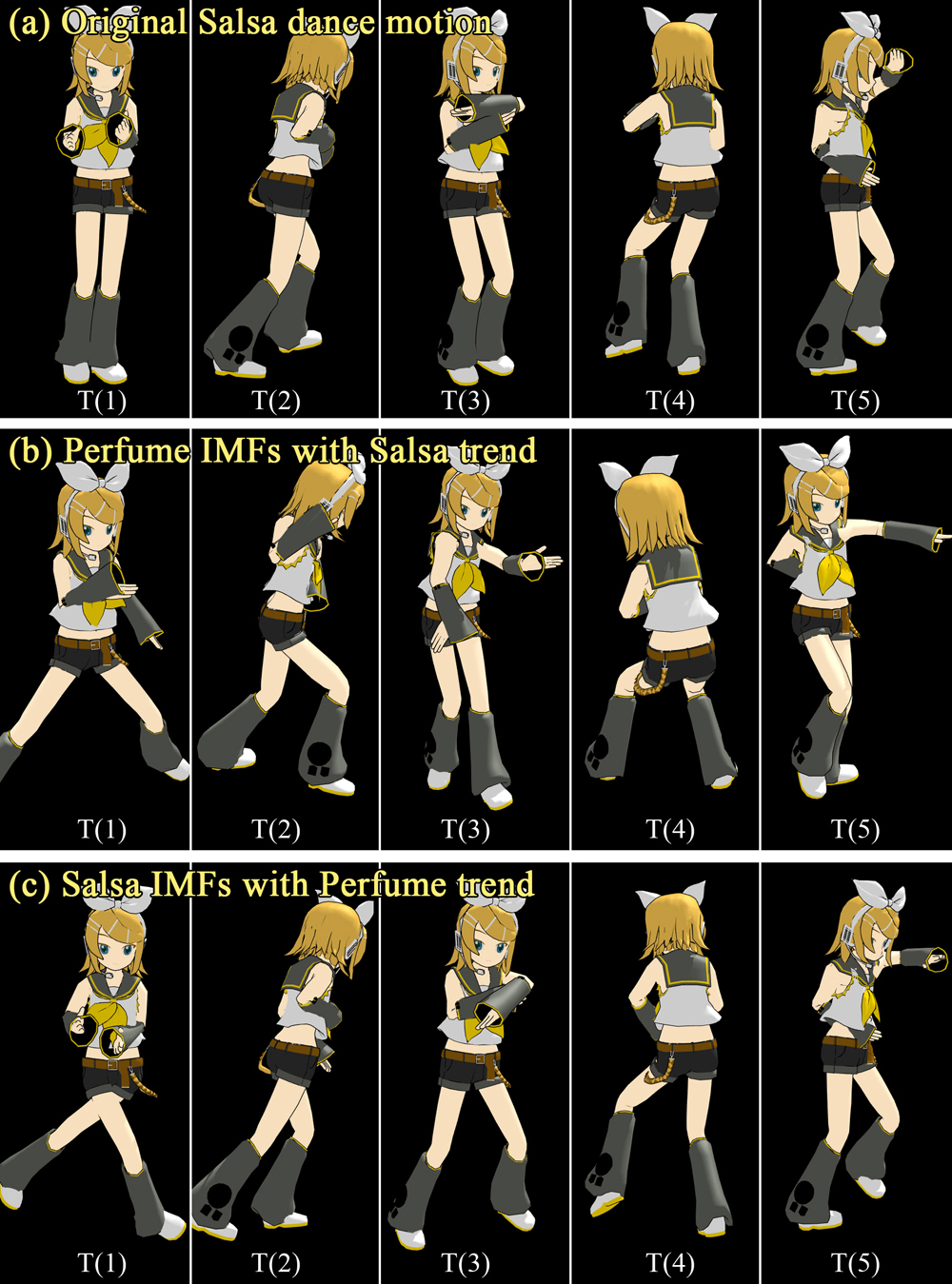}
	\caption{(a) Original Salsa dance motion and (b) Perfume dance motion with Salsa trend (c) Salsa dance motion with Perfume trend.}
\end{figure}

First, we decompose threesome pop unit Perfumes' dance into IMFs, select one of three dancers' IMFs, and generate a new dance using Salsa trend as shown in Fig. 10 (b).   For instance, as shown in Fig. 10 (b), the dancer has the entirely different posture comparing with the original one as illustrated in T (3) of Fig. 10 (a).  As a result, a new dance motion has been created.  The new dance has both Perfumes' distinct primitive motions (puppet-like choreography) and active and flexible posture transitions of the Salsa as indicated in Fig. 10 (b).  

Second, a new dance is created combining one Perfumes' trend and Salsa IMFs as shown in Fig. 10 (c). Significant differences can be observed between the original and newly created dance motions. For instance, it is clear that distinct leg motions can be observed in T (1) due to the Perfume trend in Fig. 9.

\section{Conclusions}

The human motions like dance motions are very noisy and extremely difficult to analyze. The Hilbert-Huang transform using the NA-MEMD can clearly decompose the noisy dance motions into distinct "monochromatic" IMFs and may have a substantial advantage over the other methods such as the short-time-Fourier-transform (STFT) and Wavelet analysis, etc. We propose a dance motion analysis system based on HHT, and we also discover some new dance motion characteristics using our method such as Fibonacci relations in the Hilbert spectrum, especially in the Perfume dance motions.

The Fibonacci relation in the averaged frequencies can be interpreted as follows: Because the human body has a dynamic link structure when the dancer moves each part of the body at a particular frequency, we expect that the relationship of a Fibonacci sequence can be observed in the Hilbert spectrum. We only discuss this relation briefly, and further studies are required in the future. Apparently, such characteristics cannot be discovered using other methods as we discussed in Section 3.  However, further researches are required for the detailed dance motion analyses using the HHT.

Because the EMD can decompose dance motions into "monochromatic" motions that roughly correspond to "motion primitives" with the different frequencies and its residual, i. e., Trend. For examples, we show the Perfume dance motions with the Salsa steps in Fig. 2 and the Perfume dance with the Salsa trend, and the Salsa dance with the Perfume trend as shown in Fig. 10.  Thus, new dance motions with different tastes can be created.  The editing and blending methods shown here are still limited and very primitive. Some feet-sliding, body collisions, and body penetrations can be observed using our method. These have to be carefully removed.  Further researches are required for the detailed dance motion editing and blending using the HHT in the future.

\begin{acks}
We would like to thank Mr. Daito Manabe providing the Perfume motion data for us.  We also would like to thank CAVElab members for various help and comments. 
	
\end{acks}

\bibliographystyle{ACM-Reference-Format}
\bibliography{template2}

\end{document}